\documentclass[aps,prl,superscriptaddress,amsmath,amssymb,floatfix,twocolumn]{revtex4-1}
\usepackage{times}
\usepackage{graphicx}
\usepackage{subfigure}
\usepackage{color}
\usepackage{epstopdf}

\newcommand{\mbm}{\mathbf{m}}
\newcommand{\mbq}{\mathbf{q}}
\newcommand{\mbQ}{\mathbf{Q}}
\newcommand{\mbk}{\mathbf{k}}

\begin{document}

\title{
Emergent phases in iron pnictides:
Double-Q antiferromagnetism,
charge order and
\\enhanced nematic correlations
}

\author{Rong Yu}
\email{rong.yu@ruc.edu.cn}
\affiliation{Department of Physics and Beijing Key Laboratory of Opto-electronic Functional Materials and Micro-nano Devices, Renmin University of China, Beijing 100872, China}
\affiliation{Department of Physics and Astronomy, Shanghai Jiao Tong
University, Shanghai 200240, China and Collaborative Innovation
Center of Advanced Microstructures, Nanjing 210093, China}
\author{Ming Yi}
\affiliation{Department of Physics, University of California Berkeley, Berkeley, CA 94720, USA}
\author{Benjamin A. Frandsen}
\affiliation{Department of Physics, University of California Berkeley, Berkeley, CA 94720, USA}
\affiliation{Materials Sciences Division, Lawrence Berkeley National Laboratory, Berkeley, California 94720, USA}
\author{Robert J. Birgeneau}
\affiliation{Department of Physics, University of California Berkeley, Berkeley, CA 94720, USA}
\affiliation{Materials Sciences Division, Lawrence Berkeley National Laboratory, Berkeley, California 94720, USA}
\affiliation{Department of Materials Science and Engineering, University of California Berkeley, Berkeley, CA 94720, USA}
\author{Qimiao Si}
\email{qmsi@rice.edu}
\affiliation{Department of Physics \& Astronomy, Rice University, Houston, Texas 77005,USA}

\begin{abstract}
Electron correlations produce a rich phase diagram in the iron pnictides. Earlier theoretical studies on the correlation effect demonstrated
how quantum fluctuations weaken and concurrently suppress a $C_2$-symmetric single-Q antiferromagnetic order and a nematic order.
Here we examine the emergent phases near the quantum phase transition. For a $C_4$-symmetric collinear double-Q antiferromagnetic order,
we show that it is accompanied by both a charge order and an {\it enhanced} nematic susceptibility.
Our results provide understanding for several intriguing recent experiments in hole-doped iron arsenides,
and bring out common physics that underlies the different magnetic phases of various iron-based superconductors.
\end{abstract}

\maketitle


{\it Introduction.~}
The understanding of iron-based superconductors (FeSCs) is still in flux, in part due to the entwining of various degrees of freedom
(spin, orbital and nematic)~\cite{Kamihara2008,Johnston,Dai2015,NatRevMat:2016, Hirschfeld2016, FWang-science2011}.
It was recognized since the beginning of the field that superconductivity in these systems
is driven by electron-electron Coulomb interactions rather than electron-phonon couplings.
An influential notion is that superconductivity develops at the border of
correlation-induced electronic orders. As such, a major focus
of the field is to explore a variety of electronic orders, and a rich phase diagram has been
uncovered
~\cite{Johnston,Dai2015,NatRevMat:2016, Hirschfeld2016, FWang-science2011}.
In parallel, the overall effect
of electron correlations
was emphasized from early on~\cite{Basov.2009,Si2008,Haule08,Yi_PRL2015,Wang_PRB2015}.

Consider the case of iron pnictides. Structurally, it
is comprised of layers of FeAs, each containing a square lattice of Fe ions.
Lowering the temperature leads to a tetragonal-to-orthorhombic distortion,
at $T_s$, which is accompanied or closely followed by a
N\'eel transition at $T_N \le T_s$. The structural transition
is driven by an electronic nematic transition \cite{IFisher.2012}.
The antiferromagnetic (AF) order occurs at the wavevector $(\pi,0)$ or $(0,\pi)$.
Such a single-Q AF order
reduces
 the $C_4$ symmetry of the square lattice down to $C_2$.
In this family of FeSCs,
the effect of electron correlations has been inferred from the bad-metal characteristics
in their normal state \cite{Johnston,Dai2015,NatRevMat:2016,Basov.2009,Si2008}.
It is further underscored by the recent
observation of Mott-insulating behavior in the Cu-doped iron pnictide~\cite{Song2017}.

An important question is how the single-Q AF order and the
nematic order evolve with tuning parameters such as chemical doping.
An early theoretical analysis~\cite{Dai_PNAS:2009,Abrahams_JPCM2011,Wu_PRB2016} demonstrated
a weakening and an eventual concurrent suppression of the two orders upon an isoelectronic
P-for-As substitution, which leads to quantum criticality.
This analysis was carried out within a Ginzburg-Landau (GL)
free energy functional, which was
derived from a $w$-expansion~\cite{Dai_PNAS:2009,Si_NJP:2009}. Here,
$w$ refers to the spectral weight of the coherent part of the single-electron excitations
near the Fermi energy. The iron pnictides are in the bad-metal regime, with $w$ being relatively small
-- on the order of $1/3$
as inferred from the observed Drude weight  and effective electron mass.
Experimental studies in CeFeAs$_{1-x}$P$_{x}$O
~\cite{delaCruzDai_PRL:2010,Luo2010}
and  BaFe$_2$As$_{2-x}$P$_x$
~\cite{jiang2009,kasahara2010,Analytis2014}
have provided ample evidence for this theoretical proposal.

As is often the case,
new phases can develop in the vicinity of quantum phase transitions.
In the hole-doped iron pnictides, a $C_4$-symmetric double-Q AF order
has been identified in (Ba,Sr)Fe$_2$As$_2$ upon (Na,K)-doping ~\cite{Avici:2014,Boehmer:2015,Allred_Osborn:2016}.
It occurs
close the quantum phase transition out of the single-Q AF and nematic orders, a regime of
optimal doping
close to the
maximal
transition temperature ($T_c$)
of superconductivity.

In this Letter, we examine the emergent phases near the
optimal doping within the GL analysis.
We point out how, for the parameter $w$
in this regime,
a competing
double-Q AF order with an accompanying $(\pi,\pi)$ charge order~\cite{Giovannetti:2011,Lorenzana:2008}
can emerge.
We find the surprising result that
this $C_4$
magnetic phase hosts strong
nematic fluctuations.
Our results
provide natural understanding of the recent experimental results in
several hole-doped 122 iron pnictides
~\cite{YiBirgeneau:2017,ShearModulus:2017,Frandsen:2017},
and lead to new insights into the universality in the magnetism across the various iron-based superconductors.
Microscopically, our approach captures the unusual dependence on $w$ in the expansion
around the electron-localization transition. More generally, our analysis
involves effective couplings between the magnetic and
charge/nematic order parameters
 and is expected to be robust given that
 they are symmetry-prescribed
~\cite{Harris1997}.

{\it Construction of the Ginzburg-Landau action.~}
We start from the $w$-expansion~\cite{Dai_PNAS:2009,Si_NJP:2009}.
The effect of the on-site Coulomb interactions
is treated by decomposing an electron operator
into a coherent part, denoted by $d_{i \alpha \sigma}^{coh}$, with a coherent weight $w$,
and an incoherent part, whose weight is $1-w$.
Here,
$i$, $\alpha$
and $\sigma$ label lattice site, orbital and spin indices;
in momentum space, the coherent electron operator is written as $d_{{\bf k} \alpha \sigma}^{coh}$.
Varying $w$ keeps track of the tuning
in the degree of electron correlations, with $w=1$ being the noninteracting limit.
Because $w$ is small in the bad-metal regime,
we will expand order-by-order in $w$.
It is convenient to introduce
a normalized electron operator,
$c_{\mbk \alpha \sigma}=(1/\sqrt{w})d_{\mbk \alpha \sigma}^{coh}$.
The prefactor ensures that this coherent $c-$electron operator has a spectral weight normalized to $1$.
Integrating out the high-energy
incoherent-electronic states gives rise to local moments,
which are labeled by ${\bf s}_{i,\alpha}$.
The effective Hamiltonian comprises $H_J$,
which describes
$J_1$ and $J_2$, the nearest-neighbor
and next-nearest-neighbor bilinear spin-exchange interactions,
and other terms
such as
the biquadratic interactions; $H_c$,
which characterizes
the coherent electrons as follows,
\begin{equation}\label{Eq:Hc}
 H_c=w\sum_{ij,\alpha\beta,\sigma} t_{ij}^{\alpha\beta}c_{i\alpha\sigma}^\dagger c_{j\beta\sigma}
 = \sum_{\mbk,b,\sigma}  \epsilon_{\mbk b }
 c_{\mbk b \sigma}^\dagger c_{\mbk b \sigma}
\end{equation}
with $b$ being a band index and $ \epsilon_{\mbk b } = w  \tilde{\epsilon}_{\mbk b }$ capturing
a bandwidth renormalization by the coherent weight $w$;
and  $H_m$, which specifies
an effective coupling between the
local moments and coherent electrons,
\begin{equation}\label{Eq:Hm}
 H_m=w\sum_{\mbk\mbq,
b b ^\prime, \alpha, \sigma\sigma^\prime}
\tilde{g}_{\mbk\mbq,
 b b^\prime, \alpha} c_{\mbk+\mbq
 b \sigma}^\dagger \frac{\boldsymbol{\tau}_{\sigma\sigma^\prime}}{2} c_{\mbk
 b^\prime \sigma^\prime} \cdot \mathbf{s}_{\mbq
 \alpha},
\end{equation}
where
 $\boldsymbol{\tau}$ describes the Pauli matrices.
Importantly, both $\tilde{\epsilon}$ and $\tilde{g}$ are of order $w^0$.
This procedure is an expansion
with respect to $w=0$,
which corresponds
to the threshold
interaction
for an electron localization.

\begin{figure}[t!]
\centering\includegraphics[
width=75mm
]{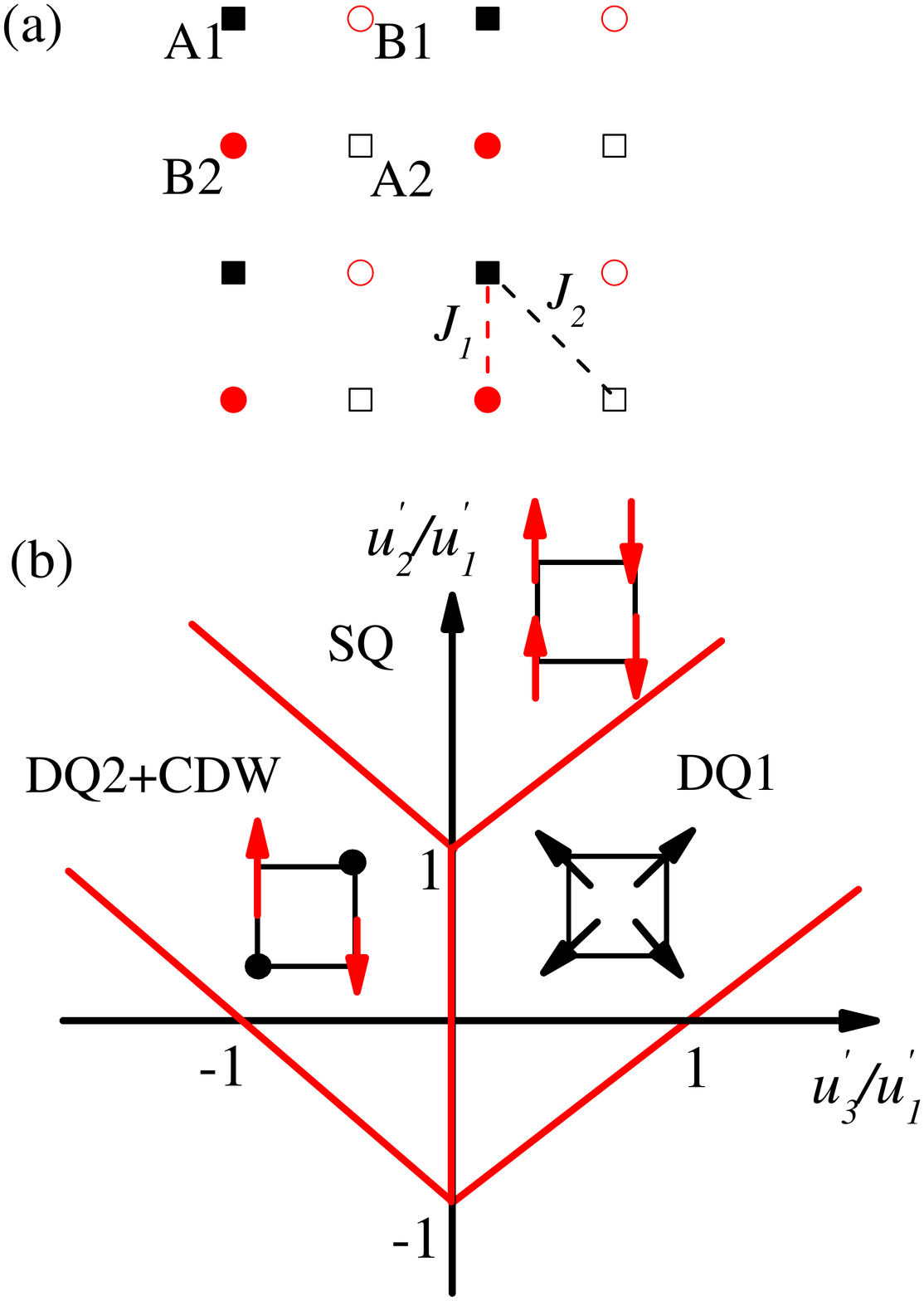}
\caption{(Color online) (a): The
Fe square lattice being divided into
four sublattices, $A1$, $A2$, $B1$, and $B2$.
(b): Ground-state phase diagram of the free-energy functional, Eqn.~\eqref{Eq:Fm2}.
The phase boundaries are marked
by the red lines.
Here, SQ, DQ1, and DQ2 refer to, respectively, the single-Q
AF phase,
the noncollinear double-Q AF phase,
and the collinear double-Q AF phase.
For each phase,
the  spin pattern on
a plaquette of the Fe square lattice is
illustrated.
}
\label{fig:1}
\end{figure}

We construct an effective action in terms of the
staggered magnetic moments, $\mbm_{A/B}$, on sublattices $A/B$.
As illustrated in Fig.~\ref{fig:1}(a), an Fe square lattice is separated into two sublattices, $A$ and $B$,
The $A$ sublattice is further
divided into $A1$ and $A2$ sublattices and, likewise, $B$ into $B1$ and $B2$.
We denote the uniform magnetization of the sublattice $A1$ by $\mbm_{\rm{A1}}$, and
similarly for the sublattices $A2$, $B1$ and $B2$.
The staggered magnetizations of the $A$ and $B$ sublattices are
\begin{align}\label{Eq:M_AB}
\mbm_{\rm{A}} &= \mbm_{\rm{A1}}-\mbm_{\rm{A2}}, \nonumber \\
\mbm_{\rm{B}} &= \mbm_{\rm{B1}}-\mbm_{\rm{B2}} .
\end{align}
To the zeroth-order in $w$,
the action is constructed from
$H_J$.
The
terms at nonzero orders in $w$ are generated by
integrating out the coherent $c$-electrons in
$H_c$ and $H_m$, both of which
are linearly proportional to $w$.
The resulting GL action, expressed in terms of $\mbm_{\rm{A/B}}$, is as follows
~\cite{Dai_PNAS:2009,Abrahams_JPCM2011}.
\begin{eqnarray}
\label{Eq:GL_Action}   S &=& S_2 +S_4, \\
\label{Eq:GL_ActionS2} S_2 &=& \sum_{\mbq,l} \left\{ \chi_0^{-1}(\mbq,i\omega_l) \left[\mbm^2_{A}(\mbq,i\omega_l) + \mbm^2_{B}(\mbq,i\omega_l)\right] \right.\nonumber\\
&& \left. + 2v(q^2_x-q^2_y) \mbm_A(\mbq,i\omega_l)\cdot\mbm_B(-\mbq,-i\omega_l)
  \right\}, \\
\label{Eq:GL_ActionS4} S_4 &=& \int_0^\beta d\tau \int d^2x \left\{ u_1(\mbm_A^2+\mbm_B^2)^2 -u_2(\mbm_A^2-\mbm_B^2)^2 \right. \nonumber\\
 && \left. -u_3(\mbm_A\cdot\mbm_B)^2 \right\} .
\end{eqnarray}

The quadratic part, $S_2$, contains
 $\chi_0^{-1} (\mbq,\omega_l) = r+ \omega_l^2+\gamma|\omega_l|+c\mbq^2$, where $\omega_l$
 is the Matsubara frequency and $c$ is the spin wave velocity. The  $v$ term describes a spin anisotropy in momentum space.
The constant term of the quadratic coefficient in $S_2$ -- {\it i.e.} the ``mass" term --
is $r=r_0 + \Delta r$. Here, $r_0$ is the zeroth-order term in $w$ produced by $H_J$, which is negative
in the ordered regime,
and $\Delta r = w A_{Q}$ is the
linear-in-$w$ contribution.
Note that $A_{Q}$ is given by the
static spin susceptibility for the coherent $c$-electrons and, thus,
must be positive (due to causality).
In other words, the effect of
$w$ is to produce a positive shift in the mass term,
which captures the physics that spin-flip interactions between the local moments and coherent electrons
weaken the AF order. The damping term $\gamma |\omega_l|$ has an upper cutoff frequency that is linear in $w$.
A quantum phase transition leading to the suppression of the single-Q AF order and the associated nematic order occurs when
$w$ reaches a threshold value such that it turns a negative $r_0$ into a positive $r$
~\cite{Dai_PNAS:2009,Abrahams_JPCM2011}.

Importantly,
the quartic coupling constants $u_i (i=1,2,3)$, appearing in $S_4$, also acquires linear-in-$w$ corrections (see
Supplementary Material~\cite{supp}).
This observation has not been made before, and it has important new consequences.
To see this, note that $H_m$ is linear in $w$. Upon integrating out the coherent $c$-electrons,
it would produce a $w^4$ factor in the corrections to the quartic-in-$\mbm$ couplings, $\Delta u_i$. However, due to
the linear-in-$w$ bandwidth of the $c$-electrons, as specified in $H_c$,
the integration over
the $c$-electron propagators produces a
singular factor $1/w^3$.
The net result is that
 $\Delta u_i$  are elevated
to the
unusual
order
of $w^4/w^3$;
 {\it i.e.}, they are
linear in $w$.
Since these terms are of the same order in $w$ as the shift to the mass $r$ term in the quadratic part,
$\Delta u_i$ will be sizable
near the quantum phase transition.
While $\Delta r$ must be positive,
 the contributions to $\Delta u_i$
are
associated with integrations of more than two
(at least four coherent) $c$-electron propagators and can, thus,
be of either sign.
This motivates an analysis of the
GL free-energy functional
 in the overall parameter space of the coupling constants. We now turn to this analysis.

{\it Magnetic phase diagram.~}
To study the magnetic ordering at wave vectors $\mbQ_1=(\pi,0)$ and (or) $\mbQ_2=(0,\pi)$,
we define
 $\mbm_1=\mbm(\pi,0)$ and $\mbm_2=\mbm(0,\pi)$.
They are connected to the staggered magnetization of the 
A,B sublattices via
 \begin{align}\label{Eq:M_12}
\mbm_{1} &= ( \mbm_{\rm{A}}-\mbm_{\rm{B}})/2 , \nonumber \\
\mbm_{2} &= (\mbm_{\rm{A}}+\mbm_{\rm{B}})/2 .
\end{align}
Rewriting the Landau
functional derived from Eqn.~\eqref{Eq:GL_Action} in terms of $\mbm_1$ and $\mbm_2$,
we have, up to the quartic order,
\begin{eqnarray}\label{Eq:Fm2}
 f_m &=& \frac{r_m}{2} (|\mbm_1|^2 + |\mbm_2|^2) + \frac{u_1^\prime}{4}  (|\mbm_1|^4 + |\mbm_2|^4) \nonumber\\
 &+& \frac{u_2^\prime+u_3^\prime}{2}
 (\mbm_1\cdot\mbm_2)^2 + \frac{u_2^\prime-u_3^\prime}{2} (\mbm_1\times\mbm_2)^2,
\end{eqnarray}
where $r_m=2
r$,
$u_1^\prime = 16u_1-4u_3$, $u_2^\prime = 16u_1-16u_2+4u_3$, and $u_3^\prime = -16u_2$.
We reiterate that
$r$
and
 $u_i$ have already
included the
corrections from
$w$.
Note also that
the damping term, which affects the behavior in the quantum critical regime
~\cite{Dai_PNAS:2009,Abrahams_JPCM2011,Wu_PRB2016}, is
 unimportant for the structure of the phase diagram we are analyzing in this section.

To proceed, we define $\theta$ to be the angle between the two
vectors $\mbm_1$ and $\mbm_2$,
and solve for the three independent variables, $|\mbm_1|$, $|\mbm_2|$, and $\theta$.
Taking the derivatives of the free energy with respect to these variables, we have
\begin{align}
 \frac{\partial f_m}{\partial |\mbm_1|} &= |\mbm_1|\left\{ r_m + u_1^\prime |\mbm_1|^2 + (u_2^\prime+u_3^\prime\cos2\theta) |\mbm_2|^2  \right\}=0, 
 \nonumber\\
 \frac{\partial f_m}{\partial |\mbm_2|} &= |\mbm_2|\left\{ r_m + u_1^\prime |\mbm_2|^2 + (u_2^\prime+u_3^\prime\cos2\theta) |\mbm_1|^2  \right\}=0, 
 \nonumber\\
 \frac{\partial f_m}{\partial \theta} &= -u_3^\prime |\mbm_1|^2 |\mbm_2|^2 \sin2\theta =0. \label{Eq:dFdtheta}
\end{align}
There are four solutions
to the
 above equations:
a) a paramagnetic (PM) state with $|\mbm_1|=|\mbm_2|=0$;
b) a single-Q AF (SQ) state with $|\mbm_1|=0$, $|\mbm_2|=\sqrt{\frac{-r_m}{u_1^\prime}}$, or $|\mbm_2|=0$,
$|\mbm_1|=\sqrt{\frac{-r_m}{u_1^\prime}}$;
c) a noncollinear double-Q AF state
(DQ1) with $\cos\theta=0$ ($\mbm_1\perp\mbm_2$) and $|\mbm_1|=|\mbm_2|=\sqrt{\frac{-r_m}{u_1^\prime+u_2^\prime-u_3^\prime}}$;
and d) a collinear double-Q AF state
(DQ2) with $\sin\theta=0$ ($\mbm_1\parallel\mbm_2$) and $|\mbm_1|=|\mbm_2|=\sqrt{\frac{-r_m}{u_1^\prime+u_2^\prime+u_3^\prime}}$.

We restrict
to the regime
where the quartic couplings are adequate to
determine the phase competition:
$u_2^\prime-u_3^\prime>-u_1^\prime$ and $u_2^\prime+u_3^\prime>-u_1^\prime$ .
The ground-state phase diagram is shown in Fig.~\ref{fig:1}(b).
In particular,
the DQ1 ground state is stabilized for $u_3^\prime>0$, $-u_1^\prime<u_2^\prime-u_3^\prime<u_1^\prime$,
and the DQ2 ground state is stabilized for $u_3^\prime<0$, $-u_1^\prime<u_2^\prime+u_3^\prime<u_1^\prime$.

We note that in the DQ2 state, the magnetic moments are ordered only at half of the iron sites (either A or B sublattice), as illustrated in the
inset of Fig.~\ref{fig:1}(b). This will have important consequences on the dynamics, to which we will return.

{\it Charge order.~}
We consider this DQ2 phase.
From the GL action in Eqns.~\eqref{Eq:GL_Action}-~\eqref{Eq:GL_ActionS4},
we perform a Hubbard-Stratonovich (HS) transformation to the $u_2$ term.
In terms of $\mbm_1$ and $\mbm_2$, the action becomes
\begin{eqnarray}
 S &=& S_2 +
 \int_0^\beta d\tau \int d^2x \left\{ 4u_1(\mbm_1^2+{\mbm_2}^2)^2 \right.\nonumber\\
 && - \left. u_3(\mbm_1^2-{\mbm_2}^2)^2 +\frac{\Delta_4^2}{u_2}+4\Delta_4(\mbm_1\cdot\mbm_2) \right\}.
\label{Eq:SmmpDelta4}
\end{eqnarray}
Here
$\Delta_4$
represents
an Ising
field
that linearly couples to $\mbm_1\cdot\mbm_2$.
It
has a
wavevector
$\mbQ=\mbQ_1+\mbQ_2=(\pi,\pi)$ and,
thus, breaks the translational symmetry of the lattice;
the $C_4$ symmetry, on the other hand, is preserved.
From the spin symmetry perspective,
since $\mbm_1\cdot\mbm_2\sim\boldsymbol{\tau}^2\sim I$,
$\Delta_4$ has the same symmetry as a charge
density
~\cite{Giovannetti:2011,Lorenzana:2008}.
Therefore, it must also linearly couple to a charge
order at wave vector $(\pi,\pi)$,
and this linear coupling locks the transition temperature of the Ising and charge ordering at $T_{co}$
~\cite{Balatsky:2010,footnote}.
With a nonzero interlayer coupling, the
DQ2 AF transition can be stabilized at
a nonzero temperature.
Since the quartic term $u_2(\mbm_1\cdot\mbm_2)^2$ is relevant  in the renormalization group sense
with respect to the temperature-induced
(classical)
AF critical point,
the thermal Ising transition can either
precede or coincide with
the magnetic one.

We close this section
by noting that
 the Ising-nematic order associated with the single-Q AF order is also seen in this
procedure by a HS transformation to the $u_3$ term of the GL action.
Here, the HS field $\Delta_2$
linearly
couples to the Ising-nematic order parameter
~\cite{FangKivelson:2008,XuMullerSachdev:2008,Dai_PNAS:2009,ChandraColemanLarkin:1990}
$\mbm_A\cdot\mbm_B$.

{\it Enhanced nematic susceptibility in the collinear double-Q antifferomagnetic phase.~}
The charge-ordered
DQ2 AF phase preserves the $C_4$ rotational symmetry.
Nonetheless, it contains the same microscopic
degrees of freedom as in the $C_2$ magnetic phase, {\it viz.}
the sublattice staggered moments
$\mbm_{A/B}$.
As noted,
the Ising-nematic field is
\begin{equation}
\label{Eq:Delta}
\Delta (x)=\mbm_A(x)\cdot\mbm_B(x) .
\end{equation}
We can thus expect considerable fluctuations of
the nematic degree of freedom.
To see this effect,
we study the
dynamical
nematic susceptibility:
\begin{eqnarray}\label{Eq:chi_nem_def}
 \chi_{\rm{nem}} (\mbq,i\nu_n) &=& 
 \int_0^\beta d\tau e^{i\nu_n\tau}
 \left\{\langle T_{\tau}
 \Delta_{\mbq,\tau} \Delta_{-\mbq,0}\rangle
  -
  \langle\Delta_{\mbq}\rangle \langle\Delta_{-\mbq}\rangle
    \right\}. 
    \nonumber\\
\end{eqnarray}
It follows from
Eqs.~(\ref{Eq:Delta},\ref{Eq:chi_nem_def})
that
$\chi_{\rm{nem}}$ involves a convolution of the fluctuations in
$\mbm_A$ and those in $\mbm_B$.
Importantly,
in the
DQ2 phase, one of the sublattices
is ordered [{\it cf.} the
 pattern in Fig.~\ref{fig:1}(b)],
giving rise to gapless magnetic excitations.
By contrast, in the paramagnetic phase,
both sublattices are disordered, and the
fluctuations
of both $\mbm_A$ and $\mbm_B$
are gapped.
We can then expect
an enhanced
$\chi_{\rm{nem}}$
in the
DQ2 AF phase, even though
there is no
 static nematic order.

To illustrate our point, we  consider the
zero-temperature
nematic susceptibility in the
DQ2 AF
phase
and compare it with
that of the paramagnetic (PM)
phase.
We do so
using a large-$N$ approach to the GL action in Eqn.~\eqref{Eq:GL_Action},
which
is generalized~\cite{supp}
from Ref. ~[\onlinecite{Wu_PRB2016}].
We extend $\mbm_{A/B}$
from $O(3)$ to $O(N)$ vectors, and scale $u_i\rightarrow u_i/N$.
We then perform HS transformations to the quartic terms of the action by introducing fields $\lambda$, $\Delta_2$
and $\Delta_4$,
where $\lambda$ refers to the mass of the propagator and,
as introduced before,
$\Delta_2$ and $\Delta_4$
are the
HS fields conjugate to the corresponding Ising order parameters.
Writing
$\mbm_{A/B}=(\sqrt{N}m_{A/B},\vec{\pi}_{A/B})$, where $m_{A/B}$ and $\vec{\pi}_{A/B}$ are the longitudinal and transverse modes,
we obtain an effective free energy functional at the leading order in $1/N$ by integrating out the $\vec{\pi}$ modes.
Differentiating the free energy with respect
to $m_{A/B}$, $\lambda$, $\Delta_2$ and $\Delta_4$
leads to a set of saddle-point equations.

\begin{figure}[t]
\centering\includegraphics[
width=80mm]{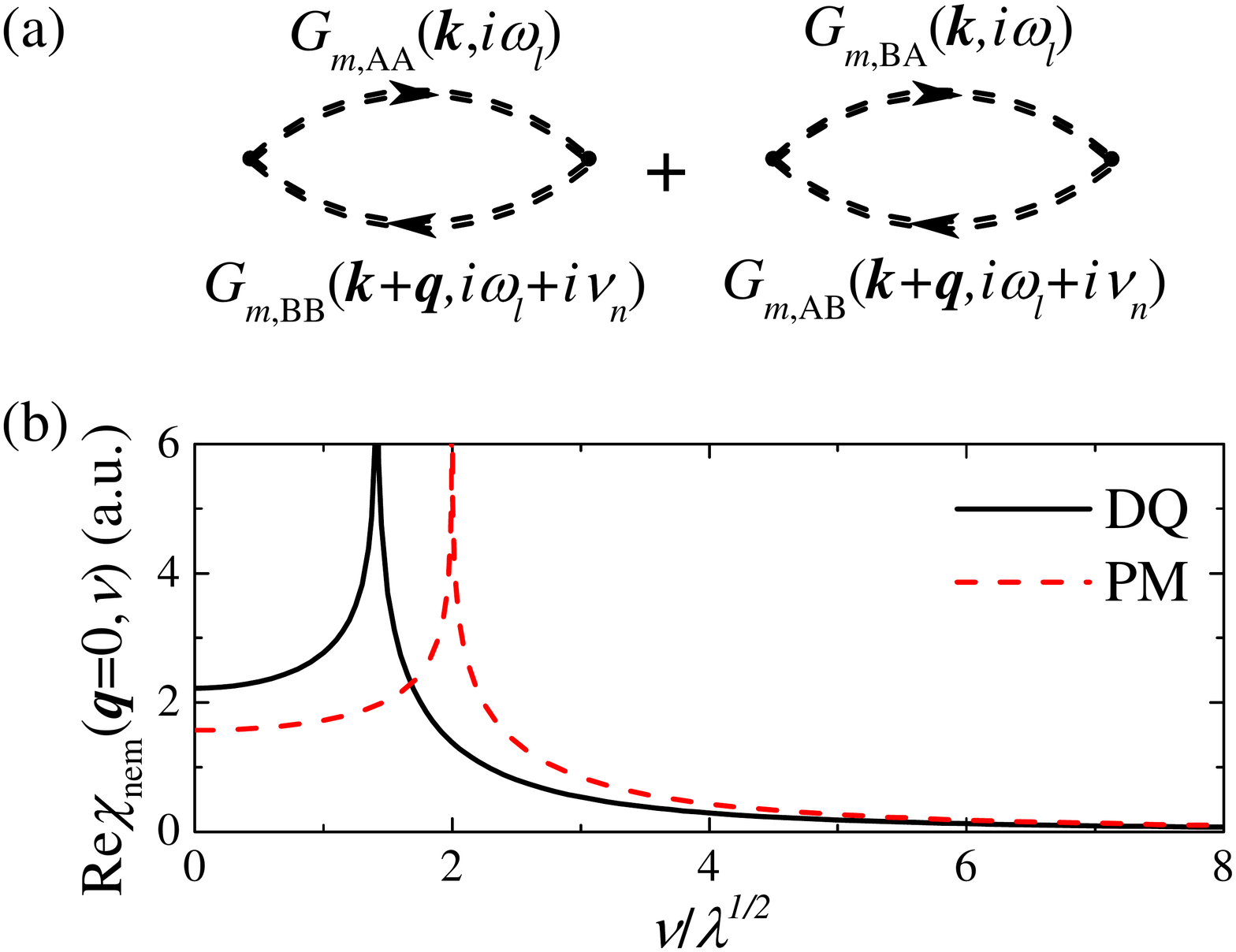}
\caption{(Color online)
(a): Feynman diagrams that contribute to the
dynamical
nematic susceptibility -- see text for details;
(b): The real part of the dynamical nematic susceptibility, $\chi_{\rm{nem}} (\mbq=0,\nu)$
in the collinear double-Q AF phase (DQ2) and the
paramagnetic phase (PM). We take the mass $\lambda=1$ in the two phases and set the spin-wave velocity $c=1$. The nematic susceptibility
is enhanced at low frequencies in the double-Q AF phase.
}
\label{fig:2}
\end{figure}

The nematic susceptibility is
 calculated at the saddle-point level in the
 DQ2 AF and
PM phases.
The difference between the nematic susceptibilities of the two cases is primarily caused by the contributions from
 the transverse magnetic modes [see Fig.~\ref{fig:2}(a) for Feynman diagrams that contribute to the nematic susceptibility].
Thus,
we will focus on
these contributions.
(Our main conclusion still holds when contributions from the longitudinal fluctuations are taken into account
~\cite{supp}).
Accordingly, the dynamical nematic susceptibility with momentum $\mbq$ and Matsubara frequency $i\nu_n$ is
\begin{eqnarray} \label{Eq:chi_nem_pi}
 && \chi_{\rm{nem}} (\mbq,i\nu_n)  \nonumber \\
  &&~~\sim T \sum_{\mbk,l} \left[ G^{\pi}_{m,AA}(\mbk,i\omega_l) G^{\pi}_{m,BB}(\mbk+\mbq,i\omega_l+i\nu_n) \right.\nonumber\\
  && ~~~~~~+ \left.  G^{\pi}_{m,BA}(\mbk,i\omega_l) G^{\pi}_{m,AB}(\mbk+\mbq,i\omega_l+i\nu_n) \right] .
\end{eqnarray}
Here,
$G^{\pi}_{m,AA} (\mbk,i\omega_l)$ {\it etc.} are
 the propagators for the
 transverse components of the
 $\mbm_{A/B}$ fields,
the details of which are
given in the Supplementary Material~\cite{supp}.

An explicit expression for the dynamical nematic susceptibility
as a function of frequency $\nu$ is obtained (see Supplementary Material~\cite{supp}).
Fig.~\ref{fig:2}(b)
shows
the real part of $\chi_{\rm{nem}}(\mbq=0,\nu)$ in the
double-Q
AF and the
paramagnetic phases.
We clearly see an enhancement of
the nematic susceptibility at low frequencies in the double-Q AF phase.
In particular, we can extract the value $\chi_{\rm{nem}}=\rm{Re}\chi_{\rm{nem}}(\mbq=0,\nu\rightarrow0)$ in each phase:
\begin{eqnarray}
 \chi_{\rm{nem}}^{\rm{PM}} 
 &\sim& \frac{\pi}{2c\sqrt{\lambda}},\\
 \chi_{\rm{nem}}^{\rm{DQ}} 
 &\sim& \frac{\pi}{c\sqrt{2\lambda}},
\end{eqnarray}
where $\lambda$ is the saddle-point value of the mass in each phase. We see that
$\chi_{\rm{nem}}^{\rm{DQ}}$
is larger than
$\chi_{\rm{nem}}^{\rm{PM}} 
$ (by a factor of $\sqrt{2}$) for the same mass value.

{\it Discussions and Conclusions.~}
Our study brings out
a key feature that connects with the experimental phase diagram
of the hole-doped iron arsenides~\cite{Allred_Osborn:2016,Boehmer:2015,Avici:2014}.
Our analysis is anchored by the reference point where
the parameters of the free energy functional are dominated by the $J_1$ and $J_2$
and related interactions between the local moments,
which place the system in the single-Q $(\pi,0)$ AF phase.
Moving towards
the optimal doping, the weight of coherent
 electrons ($w$) will increase.
At the same time that this weakens the single-Q AF order,
by reducing the magnitude of the quadratic coefficient of the free energy functional,
$r$, from the negative side,
it also changes the quartic coefficients of the free energy functional.
The latter allows for the collinear double-Q AF phase
and the accompanying charge order.

Our results on the nematic correlations in the collinear double-Q antiferromagnetic phase is consistent
with the surprising observation,
via x-ray and neutron-scattering measurements,
 of a robust local orthorhombicity~\cite{Frandsen:2017}
in the
tetragonal AF phase of Sr$_{1-x}$Na$_x$Fe$_2$As$_2$.
Moreover, our finding of
an enhanced nematic susceptibility
accompanying a charge order
in such a phase
provides a natural understanding of the
intriguing recent experiments on Ba$_{1-x}$Na$_x$Fe$_2$As$_2$,
with the ARPES studies~\cite{YiBirgeneau:2017}
and
shear-modulus measurements~\cite{ShearModulus:2017}
respectively providing evidence for a $(\pi,\pi)$ charge order and an enhanced nematic susceptibility in
its tetragonal antiferromagnetic phase.
Because the observed properties appear in the regime of optimal superconductivity,
the theoretical results advanced here will likely be important for the understanding
of superconductivity in the iron pnictides.

The observation of the $C_4$ magnetic phase lacking any nematic order seems to contrast the high-$T_c$ (Na,K)-doped (Ba,Sr)Fe$_2$As$_2$
with other iron pnictides. Our conclusion that this phase nonetheless displays strong nematic fluctuations reveals
a new connection between all these systems. Thus, our findings point to the unifying properties and common underlying physics
of the electronic orders in the iron-based superconductors.

\acknowledgements
We thank E. Abrahams, A. Frano, C. Meingast and J. Wu for useful discussions. This work has in part been supported by
the National Science Foundation of China Grant numbers 11374361 and 11674392
and Ministry of Science and Technology of China,
National Program on Key Research Project Grant number 2016YFA0300504 (R.Y.),
and by
the NSF Grant No.\ DMR-1611392, the Robert A.\ Welch Foundation Grant No.\ C-1411
and
a QuantEmX grant from ICAM and the Gordon and Betty Moore Foundation through Grant No. GBMF5305 (Q.S.),
and by the Office of Science, Office of Basic Energy
Sciences (BES), Materials Sciences and Engineering
Division, of the US Department of Energy (DOE) under
Contract No. DE-AC02-05-CH11231 within the Quantum
Materials Program (KC2202) and BES, US DOE,
Grant No. DE-AC03-76SF008 (M.Y., B.A.F., and R.J.B).
R.Y. acknowledges the hospitality of Rice University.
Q.S. acknowledges the hospitality of University of California at Berkeley
and of the Aspen Center for Physics (NSF grant No. PHY-1607611).

\newpage
\setcounter{figure}{0}
\makeatletter
\renewcommand{\thefigure}{S\@arabic\c@figure}
\onecolumngrid
\section{SUPPLEMENTARY MATERIAL --
Emergent phases in iron pnictides:
Double-Q antiferromagnetism,
charge order and enhanced nematic correlations
}
\subsection{Ginzburg-Landau action}

In this section, we further describe the construction of the Ginzburg-Landau
action
within a $w$-expansion~\cite{SMDai_PNAS:2009,SMSi_NJP:2009}.
As described in the main text,
we introduce a normalized coherent-electron operator,
$c_{{\bf k} \alpha \sigma}=(1/\sqrt{w})d_{{\bf k} \alpha \sigma}^{coh}$.
Integrating out the incoherent electrons gives rise to quasi-local moments,
${\bf s}_{i,\alpha}$, where $i$ and $\alpha$ are the site and orbital indices.
The effective local-energy Hamiltonian reads:
\begin{equation}\label{Eq:H-eff}
 H_{\rm eff} = H_J + H_c + H_m . \tag{S1}
\end{equation}
Here, $H_J$ is of zeroth order in $w$, describing an effective spin Hamiltonian for the local moments,
${\bf s}_{i,\alpha}$. Both $H_c$ and $H_m$ are linear in $w$.
\begin{align}
H_c &= \sum_{{\bf k},b,\sigma} \epsilon_{{\bf k}b\sigma}
c_{{\bf k} b \sigma}^{\dagger}
c_{{\bf k} b \sigma}
= w \sum_{{\bf k},b,\sigma} \tilde{\epsilon}_{{\bf k}b\sigma}
c_{{\bf k} b \sigma}^{\dagger}
c_{{\bf k} b \sigma}
\nonumber\\
H_m &= \sum_{{\bf k} {\bf q} b b^\prime \alpha} g_{{\bf k}{\bf
q} b b^\prime \alpha}~ c_{{\bf k}+{\bf q} b\sigma}^{\dagger}
\frac{\boldsymbol\tau_{\sigma\sigma^{\prime}}}{2}
c_{{\bf k}b^\prime\sigma^{\prime}} \cdot {\bf s}_{\mathbf{q}\alpha} \nonumber\\
&= w \sum_{{\bf k} {\bf q} b b^\prime \alpha} \tilde{g}_{{\bf k}{\bf
q} b b^\prime \alpha}~ c_{{\bf k}+{\bf q}b\sigma}^{\dagger}
\frac{\boldsymbol\tau_{\sigma\sigma^{\prime}}}{2} c_{{\bf
k}b^\prime\sigma^{\prime}} \cdot {\bf s}_{\mathbf{q}\alpha} \;. \tag{S2}
\label{H0_Hcouple}
\end{align}
Here $b$, $b^\prime$ refer to band indices, and the vector $\boldsymbol\tau$ labels the three Pauli matrices.
In the first equation, the normalized dispersion $\tilde{\epsilon}_{{\bf k}b\sigma}$ is of order $w^0$, corresponding to
 the conduction-electron dispersion at $w=1$.
The relationship
 $\epsilon_{{\bf k} b\sigma}
= w \tilde{\epsilon}_{{\bf k}b\sigma}$ reflects the band-narrowing of the coherent electrons.
The
normalized coupling
$\tilde{g}_{{\bf k}{\bf q}bb^\prime\alpha}$ is of order $w^0$.

Our goal is to construct a Ginzburg-Landau free energy functional in terms of the staggered magnetic moments on $A/B$ sublattices, $\mbm_A$ and $\mbm_B$. This can be done by integrating out the coherent $c$-electrons, and the resulting Ginzburg-Landau action is given in Eqns.~(4)-(6)
of the main text.
To the zeroth order in $w$, this effective action
is derived from $H_J$, and includes quadratic and various quartic couplings.
We will label the quadratic coefficient to be $r_0$, which is negative signifying that the spin Hamiltonian itself gives rise
to the single-Q $(\pi,0)$ [or $(0,\pi)$] AF order.

To the nonzero orders in $w$, the coupling of the local moments to the coherent electrons will modify the effective
action. In particular, it shifts
the coupling constants of both the quadratic and quartic terms, as discussed in the main text.
To see this, we define the single-particle Green's function for the $c$-electrons,
$G_{{\bf k}b}(i\omega_n)$,
and the $p$-particle Green's function representing the convolution of $p$ single-particle Green's functions
\begin{align}
\Pi^{(p)}_{{\bf q}_1\ldots{\bf q}_p}(i\omega_{n_1},\ldots,i\omega_{n_p}) &= \sum_{
i\nu_{n_1},\ldots,i\nu_{n_p}}\sum_{
{\bf k}_1,\dots,{\bf k}_p}
\sum_{\alpha_1,\ldots,\alpha_p}
\prod_{j=1}^p G_{{\bf
k}_j b_j}(i\nu_{n_j}) g_{{\bf k}_j{\bf k}_{j+1},b_jb_{j+1}\alpha_j} a_{\alpha_j} \delta({\bf k}_j+{\bf q}_j-{\bf k}_{j+1}) \delta(\nu_{n_j}+\omega_{n_j}-\nu_{n_{j+1}})
,\nonumber\\
\tag{S3}
\end{align}
where the $\delta$-functions refer to the momentum and energy conservation at each of the $p$ interacting vertices. In terms of the multi-particle Green's function, the changes in the quadratic term come from
\begin{align}
-\Pi^{(2)}_{{\bf q},-{\bf q}}(i\omega_n)
&=
\sum_{{\bf k},bb^\prime,\alpha}
g_{{\bf k}{\bf q}b b^\prime\alpha}^2
a_{\alpha}^2
\frac{ f(\epsilon_{{\bf k}+{\bf
q},b}) -f(\epsilon_{{\bf k},b^\prime})}
{i\omega_n -(\epsilon_{{\bf k}+{\bf q},b}
- \epsilon_{{\bf k},b^\prime})}
~. \tag{S4}
\label{Eq:quadratic}
\end{align}
Here,
$f(\epsilon)$ is the Fermi-Dirac distribution function
and $a_{\gamma}$ is an orbital-dependent coefficient:
$\sum_{\gamma} a_{\gamma} {\bf s}_{\gamma}$ appears
in the magnetic order-parameter ($\mbm$) field.
Note that both $g_{{\bf k}{\bf q}bb^\prime\alpha}$
and $\epsilon_{{\bf k},b^\prime}, \epsilon_{{\bf k}+{\bf q},b}$
are linear order in $w$.
Two terms follow from Eq.~(\ref{Eq:quadratic}). One is the
damping term. Generically, it has the form
$\Gamma = \gamma |\omega_n| $,
for $|\omega_n| \ll w D$ (where $D$ is the bare bandwidth, of order $w^0$).
For $\gamma$, the order in $w$ can be seen as follows: it captures a factor $w^2$ from
$g^2 \sim w^2 \tilde{g}^2$ and, at the same time, it acquires a factor $\frac{1}{w^2}$ from the
imaginary part of the retarded form $\Pi^{(2)}$. In other words,
$\gamma$ is given, to
the leading nonvanishing order in $w$,
by that
associated with the
couplings
and density of states of the $w=1$ case.
The other is a shift to $r_0$,
$\Delta r = w A_{\bf Q}$,
with
$A_{\bf Q} =
\sum_{{\bf k},b,b^\prime\alpha}
\tilde{g}_{{\bf k}{\bf q}b b^\prime\alpha}^2
a_{\gamma}^2 [\Theta(E_F-\tilde{\epsilon}_{{\bf k}+{\bf Q}})
- \Theta(E_F-\tilde{\epsilon}_{{\bf k}})]/(\tilde{\epsilon}_{{\bf k},b^\prime}
-\tilde{\epsilon}_{{\bf k}+{\bf Q},b})$
(where $\Theta$ is the Heaviside function).
Causality dictates that this shift is positive and, thus, weakens the single-Q AF order.
We stress that, naively, $\Delta r$ involves two factors of the coupling constant $g$, which would have been of order $w^2$.
However, the susceptibility, involving the convolution of two Green's functions,
scales as $\frac{1}{\epsilon } \sim \frac{1}{w} \frac{1}{\tilde{\epsilon}}$. Thus, $\Delta r$ scales as
$w^2/w$; {\it i.e.}, it is linear in $w$.

Similarly,
the changes to the quartic terms come from
\begin{align}
\Pi^{(4)}_{{\bf q}_1,{\bf q}_2,-{\bf q}_1,-{\bf q}_2}(i\omega_n,i\nu_l,-i\omega_n,-i\nu_l) + \Pi^{(4)}_{{\bf q}_1,{\bf q}_2,-{\bf q}_2,-{\bf q}_1}(i\omega_n,i\nu_l,-i\nu_l,-i\omega_n)
~. \tag{S5}
\label{Eq:quartic}
\end{align}
Since the convolution of four Green's functions for the $c$-electrons may be either positive or negative,
the corresponding changes to $u_i$ ($i=1,2,3$) could have either sign. Note that $\Pi^{(4)}\sim g^4$, which is of order $w^4$.
At the same time,
because it involves one independent integration over frequency and four Green's functions in the integrand,
 it will involve three factors
of $\frac{1}{\epsilon } \sim \frac{1}{w} \frac{1}{\tilde{\epsilon}}$.
The net result is that the leading non-vanishing order of all these corrections to $u_i$ will scale as $w^4/w^3$; {\it i.e.}, they are also linear in $w$.

The fact that both $\Delta r$ and $\Delta u_i$ are linear in $w$ seems to be surprising. They reflect the singular dependence in $w$ of the $c$-electron
correlation functions, with the two-particle susceptibility being $\sim \frac{1}{w}$ and the four-particle susceptibility susceptibility being $\sim \frac{1}{w^3}$. These singularities reflect the singular limit to the $c-$electrons
as the system approaches the electron-localization transition upon $w \rightarrow 0^{+}$.

An alternative way of seeing this is to recognize that $w$ can be factored out in $H_c + H_m$. Integrating out the coherent $c$-electrons will yield a term with $w$ multiplying a determinant. The latter can be expanded for quadratic, quartic and higher order terms in the effective Ginzburg-Landau functional. This implies that linear-in-$w$ terms appear in the quadratic, quartic and all the higher order corrections to the GL action. For the corrections to
the quadratic and quartic couplings, this is the same conclusion as above.

\subsection{General formulation of the large-$N$ approach}
Here we outline the large-$N$ approach to the Ginzburg-Landau action $S\{\mbm_A,\mbm_B\}$ in Eqns.~(4)-(6) of the main text,
which is generalized from Ref.~\cite{SMWu_PRB2016}.
The partition function and the free energy density read as
\begin{align}
 Z &= \int \mathcal{D}\mbm_A \mathcal{D}\mbm_B e^{-S\{\mbm_A,\mbm_B\}}, \tag{S6}\\
 f &= -\frac{1}{\beta V} \ln Z. \tag{S7}
\end{align}

In the large-$N$ treatment, we first generalize the sublattice magnetizations $\mbm_{A,B}$ to $O(N)$ vectors, and scale $u_i\rightarrow u_i/N$.
We
focus on the regime with all $u_i>0$,
corresponding to the case in which
the ground state is either the single-Q (SQ) AF or the collinear double-Q (DQ2) AF.
Denoting $\int dx = \int d\tau \int d^2x$, we introduce the following Hubbard-Stratonovich transformations to the $S_4$ term,
\begin{align}
 e^{-r\sum_{\mbq,l}\left[ \mbm^2_A(\mbq,\omega_l)+\mbm^2_B\right]-\int dx \frac{u_1}{N}(\mbm_A^2+\mbm_B^2)^2} &= L_1\int\mathcal{D}\lambda e^{\int dx \left\{ i\lambda (\mbm_A^2+\mbm_B^2) + \frac{N}{4u_1}(i\lambda-r)^2\right\}}, \tag{S8}\\
 e^{\frac{u_2}{N}\int dx(\mbm_A^2-\mbm_B^2)^2} &= L_2\int \mathcal{D}\Delta_4 e^{\int dx\left\{ -\frac{N\Delta_4^2}{4u_2}-\Delta_4(\mbm_A^2-\mbm_B^2)\right\}}, \tag{S9}\\
 e^{\frac{u_3}{N}\int dx(\mbm_A\cdot\mbm_B)^2} &= L_3\int \mathcal{D}\Delta_2 e^{\int dx\left\{ -\frac{N\Delta_2^2}{u_3}-2\Delta_2\mbm_A\cdot\mbm_B\right\}}, \tag{S10}
\end{align}
where the constants
\begin{equation}
 L_i=\prod_x \sqrt{\frac{u_i}{N\pi}}, \tag{S11}
\end{equation}
for $i=1,2,3$, respectively.

The partition function is then written as
\begin{align}
 Z &= L_1 L_2 L_3 \int \mathcal{D}[\mbm_A,\mbm_B,\Delta_2,\Delta_4,\lambda] \exp\{-S([\mbm_A,\mbm_B,\Delta_2,\Delta_4,\lambda])\},\nonumber\\
 &= L_1 L_2 L_3 \int \mathcal{D}[\mbm_A,\mbm_B,\Delta_2,\Delta_4,\lambda] \exp\left\{ \int dx \left[ -\frac{N\Delta_2^2}{u_3} -2\Delta_2\mbm_A\cdot\mbm_B -\frac{N\Delta_4^2}{4u_2} -\Delta_4(\mbm_A^2-\mbm_B^2) \right.\right. \nonumber\\
 & \left.\left. -i\lambda(\mbm_A^2+\mbm_B^2) +\frac{N}{4u_1}(i\lambda-r)^2 \right] -\sum_{\mbq,l} \left[
 \bar{\chi}_0^{-1} (\mbm_A^2+\mbm_B^2) +2v(q_x^2-q_y^2)\mbm_A\cdot\mbm_B \right]
 \right\}, \tag{S12}
\end{align}
where
\begin{equation}
 \bar{\chi}_0^{-1} = \chi_0^{-1}-r. \tag{S13}
\end{equation}

We express
$\mbm_{A/B} = (\sqrt{N}m_{A/B},\vec{\pi}_{A/B})$, where $m_{A/B}$ is the ordered (longitudinal) part and $\vec{\pi}_{A/B}$ refers to the $N-1$ transverse fluctuating modes. The partition function
is then rewritten
as  the product of the longitudinal and transverse parts:
$Z=Z_\sigma Z_{\pi}$, with the longitudinal part
\begin{align}
\label{Eq:Zm} Z_\sigma &= \int \mathcal{D}[m_A,m_B,\Delta_2,\Delta_4,\lambda] \exp{\left\{ -N\sum_{\mbq,l} \left[
 \bar{\chi}_0^{-1}(m_A^2+m_B^2) + 2v(q_x^2-q_y^2)m_A m_B \right]
 \right.} \nonumber\\
 & \left.-N\int dx \left[ 2\Delta_2 m_A m_B + i\lambda(m_A^2+m_B^2) +\Delta_4(m_A^2-m_B^2) + \frac{\Delta_2^2}{u_3} +\frac{\Delta_4^2}{4u_2} -\frac{(i\lambda-r)^2}{4u_1}\right]\right\}, \tag{S14}
\end{align}
and the transverse part
\begin{equation}
 Z_{\pi} = L_1L_2L_3 \int \mathcal{D}[\vec{\pi}_A,\vec{\pi}_B] \exp{\left\{ -\sum_{\mbq,l} \left(\vec{\pi}_A(-\mbq,-i\omega_l),\vec{\pi}_B(-\mbq,-i\omega_l)\right) G^{-1}_{m,\pi} \left(
 \begin{matrix}
 \vec{\pi}_A(\mbq,i\omega_l)\\
 \vec{\pi}_B(\mbq,i\omega_l)
 \end{matrix}\right)
 \right\}}, \tag{S15}
\end{equation}
where
\begin{equation}\label{Eq:Ginv}
 G^{-1}_{m,\pi}(\mbq,i\omega_l) = \left(
 \begin{matrix}
  \bar{\chi}_0^{-1} + i\lambda +\Delta_4 & v(q_x^2-q_y^2) +
  \Delta_2\\
  v(q_x^2-q_y^2) +
  \Delta_2 & \bar{\chi}_0^{-1} + i\lambda -\Delta_4
 \end{matrix}\right). \tag{S16}
\end{equation}
To the leading order in $1/N$, we can obtain the saddle-point solution by taking the $\mbq=0$ and $\omega_l=0$ components of the fields $m_A$, $m_B$ ,$\Delta_2$, $\Delta_4$, and $\lambda$ and integrating out the $\vec{\pi}_{A/B}$ modes. The transverse partition function is then
\begin{equation}
 Z_\pi = L_1L_2L_3 \prod_{\mbq,l} \sqrt{\pi/\det\left( G^{-1}_{m,\pi} \right)}, \tag{S17}
\end{equation}
and we reach the following free energy density:
\begin{align}\label{Eq:fden}
 f[m_A,m_B,\Delta_2,\Delta_4,\lambda] &= \frac{\Delta_2^2}{u_3} + \frac{\Delta_4^2}{4u_2} - \frac{(\lambda-r)^2}{4u_1} + 2\Delta_2 m_A m_B + \lambda (m_A^2+m_B^2) + \Delta_4 (m_A^2-m_B^2) \nonumber\\
 & + \frac{1}{2\beta V} \sum_{\mbq,l} \ln\left\{ \left(\bar{\chi}_0^{-1}+\lambda+\Delta_4 \right) \left(\bar{\chi}_0^{-1}+\lambda-\Delta_4 \right) -\left[v(q_x^2-q_y^2)+
 \Delta_2 \right]
 ^2 \right\}. \tag{S18}
\end{align}
Note that we have redefined $i\lambda\rightarrow\lambda$, where $\lambda$ is real, in Eqn.~\eqref{Eq:fden}.

By taking the derivatives of the free energy density in Eqn.~\eqref{Eq:fden} with respect to $m_{A/B}$, $\Delta_2$, $\Delta_4$, and $\lambda$, we obtain the following saddle-point equations:
\begin{align}
& \frac{\partial f}{\partial m_A} = 2 \left[ (\lambda+\Delta_4)m_A +\Delta_2 m_B \right] = 0, \tag{S19}\\
& \frac{\partial f}{\partial m_B} = 2 \left[ (\lambda-\Delta_4)m_B +\Delta_2 m_A \right] = 0, \tag{S20}\\
& \frac{\partial f}{\partial \Delta_2} = \frac{2\Delta_2}{u_3} +2m_Am_B -\frac{1}{2\beta V} \sum_{\mbq,l} \frac{2\left[ v(q_x^2-q_y^2)+
\Delta_2 \right]
}{\left( \bar{\chi}_0^{-1}+\lambda+\Delta_4 \right) \left( \bar{\chi}_0^{-1}+\lambda-\Delta_4 \right) -\left[ v(q_x^2-q_y^2)+
\Delta_2 \right]
^2} =0, \tag{S21}\\
& \frac{\partial f}{\partial \Delta_4} = \frac{\Delta_4}{2u_2} +(m_A^2-m_B^2) -\frac{1}{2\beta V} \sum_{\mbq,l} \frac{2\Delta_4}{\left( \bar{\chi}_0^{-1}+\lambda+\Delta_4 \right) \left( \bar{\chi}_0^{-1}+\lambda-\Delta_4 \right) -\left[ v(q_x^2-q_y^2)+
\Delta_2 \right]
^2} =0, \tag{S22}\\
& \frac{\partial f}{\partial \lambda} = -\frac{\lambda-r}{2u_1} +(m_A^2+m_B^2) +\frac{1}{2\beta V} \sum_{\mbq,l} \frac{2(\bar{\chi}_0^{-1}+\lambda)}{\left( \bar{\chi}_0^{-1}+\lambda+\Delta_4 \right) \left( \bar{\chi}_0^{-1}+\lambda-\Delta_4 \right) -\left[ v(q_x^2-q_y^2)+
\Delta_2 \right]
^2} =0. \tag{S23}
\end{align}

\subsection{Nematic susceptibility}
\subsubsection{Contribution to the nematic susceptibility from the transverse modes}
We now turn to calculating the zero-temperature nematic susceptibility in the $C_4$ symmetric
DQ2 phase and
its paramagnetic counterpart.
The expression
for the nematic susceptibility
is given in Eqn.~(13) of the main text.
The propagators of the $\vec{\pi}$ modes are
\begin{equation}
 G_{m,\pi} = \left(
 \begin{matrix}
  G^{\pi}_{m,AA} & G^{\pi}_{m,AB}\\
  G^{\pi}_{m,BA} & G^{\pi}_{m,BB}
 \end{matrix}\right) = \frac{1}{(\bar{\chi}_0^{-1}+\lambda)^2-\left[(v(q_x^2-q_y^2)+\Delta_2)^2+\Delta_4^2\right]} \left(
 \begin{matrix}
  \bar{\chi}_0^{-1}+\lambda-\Delta_4 & -v(q_x^2-q_y^2)
  -\Delta_2\\
  -v(q_x^2-q_y^2)
  -\Delta_2 & \bar{\chi}_0^{-1}+\lambda+\Delta_4\\
 \end{matrix}\right). \tag{S24}
\end{equation}

The general expression for the dynamical nematic susceptibility $\chi_{\rm{nem}}$ is
complicated, and can only be calculated numerically.
However, we can already see a clear enhancement of $\chi_{\rm{nem}} (\mbq=0,\nu\rightarrow0)$
in the
DQ2 phase
in the limit of $\gamma^2\ll\lambda$ and $v\ll c$, where we can derive analytical results.
We note that when $v\ll c$, $|G^{\pi}_{m,AB/BA}|\ll |G^{\pi}_{m,AA/BB}|$ in both the paramagnetic and the collinear double-Q AF phases.
To the leading order in $v/c$,
$\chi_{\rm{nem}} (\mbq,i\nu_n) \approx T \sum_{\mbk,l}  G^{\pi}_{m,AA}(\mbk,i\omega_l) G^{\pi}_{m,BB}(\mbk+\mbq,i\omega_l+i\nu_n) $.
In the following we show the results to
this leading order, and
consider the case $\gamma\rightarrow0$. 

In the paramagnetic phase, we define $c_{\mbk}^2=c\mbk^2+\lambda_0$, where $\lambda_0$ is the saddle-point value of the mass. At $T\rightarrow0$, we can
covert the summation over Matsubara frequencies
an integral
\begin{equation}
 T \sum_{\omega_l} \rightarrow \int_{-\infty}^\infty \frac{d\omega_l}{2\pi}. \tag{S25}
\end{equation}
We
find that
\begin{align}
 \chi_{\rm{nem}}(\mbq=0,\nu_n) &\sim \frac{N-1}{2\pi} \sum_{\mbk} \int_{-\infty}^\infty \frac{d\omega_l}{(\omega_l^2+c_{\mbk}^2)((\omega_l+\nu_n)^2+c_{\mbk}^2)}\nonumber\\
 &\sim (N-1)\sum_\mbk \frac{1}{c_{\mbk}(\nu_n^2+4c_{\mbk}^2)}\nonumber\\
 &\sim (N-1)\pi\int \frac{dk^2}{\sqrt{ck^2+\lambda_0}[\nu_n^2+4(ck^2+\lambda_0)]}\nonumber\\
 &\sim \frac{(N-1)\pi}{c\nu_n} \left[ \arctan\left( \frac{\nu_n}{2\sqrt{\lambda_0}} \right) -\arctan\left( \frac{\nu_n}{2\sqrt{c\Lambda^2+\lambda_0}} \right) \right], \tag{S26}
\end{align}
where $\Lambda$ is a momentum cutoff. Relaxing $\Lambda\rightarrow\infty$, we obtain
\begin{align}
 \chi^{\rm{PM}}_{\rm{nem}}(\mbq=0,\nu_n) &\sim \frac{(N-1)\pi}{c\nu_n} \arctan\left( \frac{\nu_n}{2\sqrt{\lambda_0}} \right) \tag{S27} 
\end{align}
Taking the analytical continuation $i\nu_n\rightarrow\nu+i0^+$ and using
 $\frac{i}{x}\arctan\left( -\frac{ix}{a} \right) = \frac{1}{2x}\ln\left( \frac{a+x}{a-x} \right)$, we
 arrive at
\begin{align}
 \rm{Re} \chi^{PM}_{\rm{nem}}(\mbq=0,\nu) &\sim \frac{(N-1)\pi}{2c\nu}\ln\left| \frac{2\sqrt{\lambda_0}+\nu}{2\sqrt{\lambda_0}-\nu} \right|. \tag{S28}
\end{align}
In the limit $\nu\rightarrow0$, we obtain Eqn.~(14) of the main text.

In the
DQ2 phase, we define
\begin{align}
 \tilde{c}_{A,\mbk}^2 &= c\mbk^2, \tag{S29}\\
 \tilde{c}_{B,\mbk}^2 &= c\mbk^2 +2\tilde{\lambda}. \tag{S30}
\end{align}
and find that
\begin{align}
 \chi_{\rm{nem}}(\mbq=0,\nu_n) &\sim \frac{N-1}{2\pi} \sum_\mbk\int_{-\infty}^\infty \frac{d\omega_l}{(\omega_l^2+\tilde{c}_{A,\mbk}^2)[(\omega_l+\nu_n)^2+\tilde{c}_{B,\mbk}^2]}\nonumber\\
 &\sim \frac{N-1}{2} \sum_{\mbk} \frac{\tilde{c}_{A,\mbk} \tilde{c}_{B,\mbk}}{\tilde{c}_{A,\mbk}+\tilde{c}_{B,\mbk}} \frac{1}{\nu_n^2+(\tilde{c}_{A,\mbk}+\tilde{c}_{B,\mbk})^2}\nonumber\\
 &\sim \frac{(N-1)\pi}{2} \int dk^2 \left( \frac{1}{\sqrt{ck^2}}+\frac{1}{\sqrt{ck^2+2\tilde{\lambda}}} \right) \frac{1}{\nu_n^2 +\left( \sqrt{ck^2}+\sqrt{ck^2+2\tilde{\lambda}} \right)^2}\nonumber\\
 &\sim \frac{(N-1)\pi}{c\nu_n} \left[ \arctan\left( \frac{\nu_n}{\sqrt{2\tilde{\lambda}}} \right) -\arctan\left( \frac{\nu_n}{\sqrt{c\Lambda^2+2\tilde{\lambda}}+\sqrt{c\Lambda^2}} \right) \right]. \tag{S31}
\end{align}
Taking the momentum cutoff $\Lambda\rightarrow\infty$, we
now yield
\begin{align}
 \chi^{\rm{DQ}}_{\rm{nem}}(\mbq=0,\nu_n) &\sim \frac{(N-1)\pi}{c\nu_n}\arctan\left( \frac{\nu_n}{\sqrt{2\tilde{\lambda}}} \right). \tag{S32} 
\end{align}
After the analytical continuation $i\nu_n\rightarrow\nu+i0^+$,
this gives rise to
\begin{align}
 \rm{Re} \chi^{DQ}_{\rm{nem}}(\mbq=0,\nu) &\sim \frac{(N-1)\pi}{2c\nu}\ln\left| \frac{\sqrt{2\tilde{\lambda}}+\nu}{\sqrt{2\tilde{\lambda}}-\nu} \right|. \tag{S33}
\end{align}
Taking the limit $\nu\rightarrow0$, we reach Eqn.~(15) of the main text.

\subsubsection{Contribution to the nematic susceptibility from the longitudinal modes}
To take into account the longitudinal fluctuations, we introduce $\sigma_{A/B}$, $\delta\lambda$, $\delta\Delta_2$, and $\delta\Delta_4$ as fluctuations of the corresponding fields $m_{A/B}$, $\lambda$, $\Delta_2$, and $\Delta_4$, and expand the longitudinal part of the partition function $Z_m$, given in Eqn.~\eqref{Eq:Zm}, up to the quadratic terms
around the saddle point. Defining the vector
\begin{align}
 \vec{\Sigma}(\mbq,i\omega_l) &= \left( \sigma_A, \sigma_B, \delta\lambda, \delta\Delta_2, \delta\Delta_4 \right)^T, \tag{S34}
\end{align}
we can express
\begin{align}
 Z_m &= Z_{m0} +  \int \mathcal{D}[\vec{\Sigma}] \exp{\left\{ -\sum_{\mbq,l} \vec{\Sigma}^T(-\mbq,-i\omega_l) G^{-1}_{m,\sigma}
 \vec{\Sigma} (\mbq,i\omega_l)
 \right\}}, \tag{S35}
\end{align}
where $Z_{m0}$ is the saddle-point partition function. Here,
$G^{-1}_{m,\sigma}$ is the inverse longitudinal propagator and takes the following matrix form:
\begin{align}
 G^{-1}_{m,\sigma} &= \left(
 \begin{matrix}
  \bar{\chi}_0^{-1}+\lambda+\Delta_4 & \Delta_2 & \sqrt{N}m_A & \sqrt{N}m_B & \sqrt{N}m_A \\
  \Delta_2 & \bar{\chi}_0^{-1}+\lambda-\Delta_4 & \sqrt{N}m_B & \sqrt{N}m_A & \sqrt{N}m_B \\
  \sqrt{N}m_A & \sqrt{N}m_B & -\frac{N}{4u_1} & 0 & 0 \\
  \sqrt{N}m_B & \sqrt{N}m_A & 0 & \frac{N}{u_3} & 0 \\
  \sqrt{N}m_A & \sqrt{N}m_B & 0 & 0 & \frac{N}{4u_2}
 \end{matrix}
 \right). \tag{S36}
\end{align}

In the paramagnetic phase, $m_A=m_B=\Delta_2=\Delta_4=0$.
In this case, the longitudinal propagators are $G_{m,AA}^\sigma=G_{m,BB}^\sigma=1/(\bar{\chi}_0^{-1}+\lambda_0)$,
and $G_{m,AB}^\sigma=G_{m,BA}^\sigma=0$, identical to the transverse ones.
This is expected as the model is $SU(2)$ symmetric in this phase. (Here, we are using the terms ``longitudinal" and ``transverse"
 in reference to its ordered counterpart.)
 We then
 find the contribution to the nematic susceptibility to be
\begin{align}
  \chi^{\rm{PM},\sigma}_{\rm{nem}}(\mbq=0,\nu_n) &\sim \frac{\pi}{c\nu_n} \arctan\left( \frac{\nu_n}{2\sqrt{\lambda_0}} \right). \tag{S37} 
\end{align}
This is also identical to the contribution from the $\vec{\pi}$ modes besides the $(N-1)$ factor. So the total
susceptibility is
\begin{align}
  \chi^{\rm{PM},tot}_{\rm{nem}}(\mbq=0,\nu_n)
  &\sim \frac{N\pi}{c\nu_n} \arctan\left( \frac{\nu_n}{2\sqrt{\lambda_0}} \right), \tag{S38} 
\end{align}
which rescales the expression in Eqn.~(28) by a factor of $N/(N-1)$.

In the double-Q phase, $\Delta_2=0$, and we take $m_B=0$. After inverting $G^{-1}_{m,\sigma}$, we find $G_{m,AB}^\sigma=G_{m,BA}^\sigma=0$, and
\begin{align}
 G_{m,AA}^\sigma &= \frac{1}{\bar{\chi}_0^{-1}+4(u_1-u_2)m_A^2}, \tag{S39} \\
 G_{m,BB}^\sigma &= \frac{1}{\bar{\chi}_0^{-1}+2\tilde{\lambda}-u_3m_A^2}. \tag{S40}
\end{align}

We introduce
\begin{align}
 (\tilde{c}^\prime_{A,\mbk})^2 &= c\mbk^2 + (u_1-u_2) m_A^2, \tag{S41}\\
 (\tilde{c}^\prime_{B,\mbk})^2 &= c\mbk^2 +2\tilde{\lambda} - u_3 m_A^2, \tag{S42}
\end{align}
in a similar way as in the treatment of
 the $\vec{\pi}$ modes, and we find the following form for the longitudinal contribution to the nematic susceptibility
\begin{align}
 \chi^{\rm{DQ},\sigma}_{\rm{nem}}(\mbq=0,\nu_n) &\sim \frac{1}{2\pi} \sum_\mbk\int_{-\infty}^\infty \frac{d\omega_l}{(\omega_l^2+(\tilde{c}_{A,\mbk}^\prime)^2)[(\omega_l+\nu_n)^2+(\tilde{c}_{B,\mbk}^\prime)^2]}\nonumber\\
 &\sim \frac{\pi}{c\nu_n} \arctan\left( \frac{\nu_n}{\sqrt{(u_1-u_2)m_A^2}+\sqrt{2\tilde{\lambda}-u_3m_A^2}} \right). \tag{S43} 
\end{align}
In general, by solving the saddle-point equations, we can show that $\sqrt{(u_1-u_2)m_A^2}+\sqrt{2\tilde{\lambda}-u_3m_A^2}<2\sqrt{\lambda_0}$, which also lead to a larger contribution to nematic susceptibility than in the paramagnetic phase. In particular, when the magnetic order is weak, \emph{i.e.}, $m_A^2\ll\tilde{\lambda}$,
this leads to
\begin{align}
 \chi^{\rm{DQ},\sigma}_{\rm{nem}}(\mbq=0,\nu_n) &\approx \frac{\pi}{c\nu_n}\arctan\left( \frac{\nu_n}{\sqrt{2\tilde{\lambda}}} \right), \tag{S44}
\end{align}
which is approximately the same
as  the transverse contribution, except for the $(N-1)$ factor.
In this regime, the total contribution is
approximately
\begin{align}
  \chi^{\rm{DQ}, tot}_{\rm{nem}}(\mbq=0,\nu_n) &\approx \frac{N\pi}{c\nu_n}\arctan\left( \frac{\nu_n}{\sqrt{2\tilde{\lambda}}} \right). \tag{S45}
\end{align}
We can now
compare Eqn.~(S45) with Eqn.~(S38), and
see that our main conclusion on the enhancement of the nematic susceptibility still holds
upon the inclusion of the contribution from longitudinal fluctuations.




\begin{thebibliography}{99}
\bibitem{footnote}
We note in passing that, in the noncollinear double-Q phase (DQ1) phase in the phase diagram of Fig.~\ref{fig:1},
$\mbm_1\cdot\mbm_2=0$. Thus, this phase is not accompanied by a charge order.
\end{thebibliography}
\end{document}